# Holography, Dark Energy and Entropy of Large Cosmic Structures

*C Sivaram*

Indian Institute of Astrophysics, Bangalore, 560 034, India

Telephone: +91-80-2553 0672; Fax: +91-80-2553 4043

e-mail: sivaram@iiap.res.in

*Kenath Arun*

Christ Junior College, Bangalore, 560 029, India

Telephone: +91-80-4012 9292; Fax: +91-80- 4012 9222

e-mail: kenath.arun@cjc.christcollege.edu

**Abstract:** As is well known, black hole entropy is proportional to the area of the horizon suggesting a holographic principle wherein all degrees of freedom contributing to the entropy reside on the surface. In this note, we point out that large scale dark energy (such as a cosmological constant) constraining cosmic structures can imply a similar situation for the entropy of a hierarchy of such objects.

The holographic principle (Susskind, 1995; 't Hooft, 1993) has been invoked in connection with the well known result that the entropy of black holes scales with the area of the horizon (rather than volume like other systems). Thus:

$$S = k_B \frac{c^3}{4G\hbar} A \qquad \ldots (1)$$

Where $A$ is the black hole horizon given by:

$$A = 4\pi \left(\frac{2GM}{c^2}\right)^2 \qquad \ldots (2)$$

M is the black hole mass.

This implies: $S \cong k_B \frac{A}{L_{Pl}^2} \qquad \ldots (3)$

Where, $L_{Pl} = \left(\frac{\hbar G}{c^3}\right)^{1/2}$ is the Planck length.

Now in recent papers (Sivaram, 1994a; 2008; Sivaram & Arun, 2012a) a new kind of cosmological paradigm was invoked wherein the requirement that for a hierarchy of large scale structures, like galaxies, galaxy clusters, super-clusters, etc. their gravitational (binding) self energy density must at least equal or exceed the background repulsive dark energy density (a cosmological constant as current observations strongly suggests) implies a mass-radius relation of the type:

$$\frac{M}{R^2} = \frac{c^2}{G}\sqrt{\Lambda} \qquad \ldots (4)$$

(The requirement that gravitational self energy density $= \frac{GM^2}{8\pi R^4}$ should be comparable to the background cosmic vacuum energy density of $\frac{\Lambda c^4}{8\pi G}$ for the object to be gravitationally bound (autonomous) structures, implies equation (4) above (Sivaram, 1994a; 2007; 2008)

$\Lambda$ here is the cosmological constant with an observed value of $10^{-56} cm^{-2}$.

Thus: $M \propto R^2 \qquad \ldots (5)$

Equations (4) and (5) are seen to hold for a whole range of large scale structures, including the Hubble volume. Thus:

$$R_{gal} = 3 \times 10^{22} cm \Rightarrow M_{gal} = 10^{45} g$$
$$R_{clus} = 3 \times 10^{24} cm \Rightarrow M_{clus} = 10^{49} g$$

This relation holds right down to globular clusters. (Sivaram, 1994b)

Considering that these structures are constituted by N particles of mass m, then the entropy can be written as:

$$S = k_B N \qquad \ldots (6)$$

(for identical particles)

Thus (from equation (4)):

$$k_B N \propto k_B \frac{M}{m} \propto k_B \frac{R^2}{m} \qquad \ldots (7)$$

Using equation (4) we have:

$$S = k_B \sqrt{\Lambda} \frac{c^2}{Gm} R^2 \qquad \ldots (8a)$$

For $m = m_P$, the proton mass, we have:

$$S = k_B \sqrt{\Lambda} \frac{c^2}{Gm_P} R^2 \qquad \ldots (8b)$$

So we see that the entropy is proportional to the surface area of the structure, given by $4\pi R^2$.

As $\Lambda$, $m_P$, $c$, $G$ are constants, the entropy just involves the area, thus implying a new holographic principle similar to that for black holes. We can estimate $S$ for different structures.

For a large galaxy, $R_{gal} \approx 10^{23} cm; S \approx 10^{68} k_B$ (corresponding to $10^{68}$ protons)

For a galactic cluster, $R_{clus} \approx 10^{25} cm; S \approx 10^{72} k_B$

This also gives the baryonic entropy $\sim 10^{78} k_B$ for the Hubble radius.

For the average energy of the CMBR photon $\sim 10^{-12} GeV$ this gives the total entropy of the radiation as $\sim 10^{88} k_B$. Thus basically, we have entropy proportional to area, suggesting some kind of cosmic holography, distinct from that for black holes. In the case of black holes, we have:

$$S_{bh} = k_B \frac{A}{L_{Pl}^2}$$

Whereas for these cosmic structures we have:

$$S = k_B \frac{\sqrt{\Lambda}}{\left(Gm_P/c^2\right)} A \qquad \ldots (9)$$

Now $\left(Gm_P/c^2\right) \approx 2 \times 10^{-52} cm$ is the proton-Schwarzschild radius. As $\left(\sqrt{\Lambda}\right)^{-1}$ gives a length scale of $\sim 10^{28} cm$, i.e. of the order of the Hubble radius, it is interesting that

$$\frac{\sqrt{\Lambda}}{\left(Gm_P/c^2\right)} \approx 10^{-24} cm^2 \sim (proton\ radius)^2 \qquad \ldots (10)$$

So equation (9) can be written as:

$$S = k_B \frac{A}{L_{Pr}^2} \qquad \ldots (11)$$

Where $L_{Pr}$ is the proton radius.

So while black hole entropy can be pictured as the total number of 'Planck areas', covering the horizon surface area (Sivaram & Arun, 2009; Bekenstein, 1973; 1975), the entropy for these cosmic structures (constrained by dark energy) is given by the number of 'proton areas' covering the surface area of these objects. If dominated by dark matter particles of mass $m_D$ and the fraction of dark matter contributing to the mass $M$ is $f$, then the entropy due to the dark matter particles is just their number, i.e. $\frac{fM}{m_D}$, which by equation (9) would still be proportional to the area of the structure. (So we can add up the entropies for the different constituent particles, i.e. $f_1$, $f_2$, etc. each being proportional to the area).

As the Planck area $L_{Pl}^2$ is much smaller than the proton radius squared, the black hole entropy would be $10^{38}$ times larger, i.e. if the galaxy mass collapses to a black hole its entropy would be $\sim 10^{105} k_B$, i.e. $10^{38}$ times larger! So if the universe were to be a black hole, its entropy would increase by the same factor, $\sim 10^{123} k_B$, a well quoted figure in the literature, as maximum entropy or information content contained within a Hubble radius.

In (Sivaram, 1994a) it was also pointed out that remarkably enough the $M/R^2$ relation as given by equations (4) and (5) also hold for individual elementary particles like a proton, i.e. $R_P \sim 10^{-12} cm \Rightarrow m_P = 10^{-24} g$. So that the entropy as given by equations (9) and (11) would just be $k_B$ (the smallest unit of entropy)! So the holographic principle could go down to the level of individual fundamental particles! This was justified much earlier in (Sivaram, 1982; Sivaram & Arun, 2012b).

Thus entropy/area is a constant for black holes; the constant being $\left(c^3/\hbar G\right) = L_{Pl}^{-2} \approx 10^{66}$. And entropy/area is also a constant for all the above large scale structures (constrained by dark energy) and the constant is now $\dfrac{\sqrt{\Lambda}}{Gm_P/c^2} \approx 10^{24}$.

$\Lambda$ plays the role of background curvature (Sivaram, 1994a, 1994b), so we can write:

$$\frac{M}{R^2} \approx \text{background curvature x superstring tension} \qquad \ldots (12)$$

While $\dfrac{c^2}{G}$ is just the tension in superstring theories fixing the gravitational constant (Sivaram, 1990). In the case of black holes, the background curvature is just the Schwarzschild radius inverse squared, i.e. $\dfrac{c^4}{G^2 M^2}$, so that with $R = R_S = \dfrac{2GM}{c^2}$, we have $\dfrac{M_{bh}}{R_S} = \dfrac{c^2}{G}$, which is the string tension.

This also explains why for a closed universe:

$$R_S = \frac{1}{\sqrt{\Lambda}}, \quad \frac{M}{R} = \frac{c^2}{G} \qquad \ldots (13)$$

The large scale structures are 'embedded' in a space of background curvature $\Lambda$ (which is the dark energy). The $\frac{M}{R^2}$ relation is suggestive of a membrane tension (or surface tension) which has the same universal value for all the large scale cosmic structures from globular cluster to the Hubble universal, this value being: (Sivaram, 1994a)

$$T = \frac{c^2}{G}\sqrt{\Lambda} \; ergs/cm^2 \qquad \ldots (14)$$

A kind of universal surface tension, suggesting the holographic picture

Indeed the energy (mass) per unit area, i.e. surface tension given by above equations, i.e. $\frac{M}{R^2} = \frac{c^2}{G}\sqrt{\Lambda}$, has the same numerical value as that in nuclear physics, like the surface tension in the nuclear liquid drop model of $\sim 10^{21} ergs/cm^2$ (Sivaram, 2005; 2007). This has consequences for the entropy of nuclear matter. This is a most intriguing fact, which would be explored in a subsequent work.

Thus the holographic concept for entropy goes well beyond black holes and encompasses many other objects, suggesting a deep underlying link connecting all scales.